\documentclass[twocolumn,a4paper,reprint,amssymb,aps,prd,groupedaddress,superscriptaddress,nofootinbib]{revtex4-2}
\pdfoutput=1
\usepackage{graphicx}
\usepackage{epsf}
\usepackage{bm}
\usepackage{amsmath}
\usepackage{amsfonts}
\usepackage{amssymb}
\usepackage{epstopdf}
\usepackage{natbib}
\usepackage{color}
\usepackage[dvipsnames]{xcolor}
\usepackage{physics}
\usepackage{bm}
\usepackage[colorlinks = true,
            linkcolor = blue,
            urlcolor = blue,
            citecolor = blue,
            anchorcolor = blue]{hyperref}
\usepackage{lipsum}
\providecommand{\U}[1]{\protect\rule{.1in}{.1in}}

\usepackage[capitalize]{cleveref}
\usepackage{soul}
\usepackage{braket}
\usepackage{enumitem}

\begin{document}

\title{Probing the Universe's Topology  through a Quantum System?}
\author{Evangelos Achilleas Paraskevas}
\email{e.paraskevas@uoi.gr}
\affiliation{Department of Physics, University of Ioannina, GR-45110, Ioannina, Greece}
\author{Leandros Perivolaropoulos}
\email{leandros@uoi.gr}
\affiliation{Department of Physics, University of Ioannina, GR-45110, Ioannina, Greece}

\begin{abstract}
The global topology of the Universe could, in principle, affect quantum systems through boundary condition constraints. We investigate this connection by analyzing how compact, flat, cosmologically inspired topologies—specifically the 3-Torus ($E_1$) and half-turn space ($E_2$)—influence the energy eigenvalues of a quantum particle in the bound state of a 3D Dirac delta potential. Using rigorous renormalization techniques, we derive the equations satisfied by the energy eigenvalues in each topology and develop a systematic method to compute spectral shifts. Our results reveal that each topology induces characteristic  deviations in the energy spectrum. In the large-$L$ limit ($L \gg g_R$), to leading order, the energy eigenvalues for both the $E_1$ and $E_2$ spaces can be written in the unified form
$E\simeq -\frac{\hbar^2}{2mg_R^2}\bigl(1 + C_\Gamma\,\frac{2g_R}{L}\,e^{-L/g_R}\bigr)$,
where the topology-dependent coefficient is
$C_\Gamma = 6$ for the $E_1$ space and $C_\Gamma = 4$ for the $E_2$ space, $g_R$ is the characteristic length scale of the quantum system, and $L$ is the side physical length of the fundamental cubic region. 
Using the three‐dimensional Dirac potential as a toy model, we show that at the current cosmic epoch ($a=1$), these topological effects are exponentially suppressed, rendering direct observation infeasible. However, such effects may become measurable in the early Universe, when the physical size of the particle horizon is comparable to the characteristic scale of the quantum system. While immediate experimental verification remains impractical, our work offers theoretical insight into how global cosmic topology might manifest in quantum bound states and may inform future studies of early‐Universe quantum phenomena. 
\end{abstract}

\maketitle

\section{\label{sec:introduction}Introduction}

Einstein’s general theory of relativity (GR), combined with cosmological observations, constrains the average local geometry of the Universe but does not fully determine the global cosmic topology.  A natural question arises: Could the global topology of the Universe imprint detectable signatures on local physical systems?
 
Topology of space may imprint characteristic signatures accessible to local measurements. For example, \citet{Floratos:2012nf} show that an \(\mathbb{R}^2\times\mathcal{S}^1\) topology modifies the Newtonian gravitational potential, and \citet{Iorio:2012ic} propose  local tests. Additional gravitational signatures of nontrivial topology have been explored in \citep{Roukema:2006yd,Vigneron:2022ear,Vigneron:2022eim}.  

Moreover, \citet{Perivolaropoulos:2017rgq} argues that a maximum measurable length—on the order of or exceeding the physical particle horizon (\(\gtrsim10^{26}\,\mathrm{m}\))—potentially arising from cosmic topology, alters quantum spectra via a Generalized Uncertainty Principle (GUP) \citep{Maggiore:1993rv,Tawfik:2015rva}.  \citet{Perivolaropoulos:2017rgq} further demonstrate that the induced shifts in the energy levels of typical quantum systems are extraordinarily suppressed—well beyond current experimental reach—due to the immense size of today’s particle horizon.  However, this suppression may not hold during the Early Universe, where quantum effects that generated the inflationary primordial fluctuation spectrum could amplify such deviations\citep{Skara:2019uzz}.

Building on this foundation, \citet{Skara:2019uzz} estimated the effects of such a GUP on the primordial fluctuation spectrum and its spectral index to derive observational bounds. Their analysis of CMB and large-scale structure data demonstrated that identifying the maximum measurable length with the current particle horizon ($\sim 10^{26}$ m) remains consistent with observations at the $2\sigma$ confidence level, suggesting that topological effects could have been more prominent in earlier cosmic epochs.

  Efforts to observationally determine cosmic topology have a long history and fall into two main approaches. The first involves searching for objects visible in multiple directions on the sky, corresponding to multiple null geodesics connecting us to their past worldlines. In general, if the topology scale were sufficiently small, we would observe repeated patterns, such as clones of galaxies or quasars, which have not been detected. These repeated patterns represent a real physical phenomenon that becomes observable when the characteristic length scale is small enough, along with the observer’s orientation relative to the lattice~\citep{1974JETP...39..196S,1983CoTPh...2.1055F,Lachieze-Rey:1995qrb,Lehoucq:1996qe,Fujii:2011ga,COMPACT:2022gbl}. 
 For example, in a universe with a 3-Torus spatial topology, multiple images of an object aligned along one of the principal directions would appear periodically in comoving proper distance, as space is tiled by copies of the fundamental domain. However, the shape of the fundamental domain is neither observable nor a physical property of the manifold~\citep{COMPACT:2022gbl}. 
  
  The second approach looks for signatures of topology in the fluctuations of the cosmic microwave background (CMB) temperature and polarization\citep{Levin:2001fg,Roukema:2006yd,Fujii:2011ga,COMPACT:2022gbl,COMPACT:2022nsu,COMPACT:2023rkp,COMPACT:2024cud,COMPACT:2024dqe,COMPACT:2024qni,COMPACT:2025gqg}. Similarly, in the CMB temperature fluctuations, matching pairs of circles in the sky would indicate the self-intersection of the last scattering surface (LSS). This occurs when the topology scale is smaller than the LSS diameter, resulting in pairs of circles with matching temperature (and polarization) patterns appearing in different regions of the sky\citep{Cornish:1996kv}(see also \cite{Levin:2001fg}). The matched-circles method is a generic approach: for a sufficiently small fundamental domain (though correlations can persist even when the topology scale is too large to produce matched circles\citep{Fabre:2013wia,COMPACT:2023rkp}), any nontrivial topology in an FLRW cosmology would predict such repeating circle patterns.
  Interestingly, the observed fluctuations in the CMB contain information about topology \citep{Starobinsky:1993yx,Stevens:1993zz}, even if the scale of the topology is larger than the diameter of the observable Universe. 
 To date, no evidence of nontrivial cosmic topology has been detected~\citep{COMPACT:2022gbl}.

The three plausible FLRW geometries (\( \mathbb{H}^3 \), \( \mathbb{R}^3 \), and \( \mathbb{S}^3 \)) are described by the convenient form of the FRW metric\citep{Wald:1984rg,Dodelson:2003ft,Hobson:2006se}:

\begin{equation}\label{frw}
ds^2 = -c^2 dt^2 + R^2(t) \left[ d\chi^2 + S^2(\chi) \left( d\theta^2 + \sin^2 \theta \, d\phi^2 \right) \right],
\end{equation}
where \( (\chi, \theta, \phi) \) are comoving coordinates and \(  S(\chi) \) is defined as:

\begin{equation}
S(\chi) =
\begin{cases}
\sin \chi & \text{if } k = 1, \\
\chi & \text{if } k = 0, \\
\sinh \chi & \text{if } k = -1.
\end{cases}
\end{equation}
These geometries describe the universal covering space\citep{Lachieze-Rey:1995qrb,Frankel:1997ec,Levin:2001fg}. If we assume that the space is simply connected, its universal cover is identical to the space itself. By construction, the local geometry of a space is indistinguishable from that of its universal covering space. However, for multiply connected space forms, the global properties of the space can differ significantly from those of its universal covering space. Therefore, there are many possible smooth three-spaces that correspond to each universal covering space. 
 Although the topological possibilities for \( \mathbb{H}^3 \) and \( \mathbb{S}^3 \) are diverse, they are not relevant here, as we focus on studying quantum systems in specific examples that are flat and compact.

 For a flat (\(k = 0\)) FLRW universe, there are exactly eighteen topologically distinct spatial manifolds \citep{Lachieze-Rey:1995qrb,Riazuelo:2003ud,COMPACT:2023rkp}. Eight of these are non‐orientable and are typically excluded from cosmological studies. The first ten—denoted \(E_1\) through \(E_{10}\)—are compact, implying a finite three-dimensional volume in the flat (\(k=0\)) FLRW metric; of these ten, six are orientable. The manifolds \(E_{11}\)–\(E_{15}\) feature two‐dimensional slices of finite area, whereas \(E_{16}\) and \(E_{17}\) are compact in only one spatial direction. Finally, the covering space \(E_{18}\) is simply connected and infinite in all directions.

The questions addressed in the present analysis are the following:
\begin{itemize}[left=0pt]
  \item How do flat, compact, cosmic-inspired topologies alter the bound-state energy eigenvalues of a quantum particle?
  
  \item To what extent can direct quantum measurements probe the global spatial topology of the universe?  
    Specifically, what deviations in energy eigenvalues arise for selected compact flat topologies (e.g. 3-Torus, Half-turn space) relative to an infinite flat universe?
  
   \item At which cosmological epochs (e.g.\ inflation, electroweak epoch, present day) would topologically induced spectral shifts become large enough to be detectable? Should theoretical models of early-universe phenomena—such as phase transitions and the spectrum of primordial fluctuations—account for nontrivial cosmic topology?
\end{itemize}

The structure of this paper is as follows:   First, in Section~\ref{sectionii} we review the notion of universal covering space. Secondly, in Section~\ref{sectioniii}, after reviewing the results presented in \citep{Cavalcanti:1998jx} by calculating the energy eigenvalues of a bound state in a 3D Dirac delta potential in $\mathbb{R}^3$ using a renormalized coupling (see also \citep{Atkinson:1975vv,Gosdzinsky:1990vz,Manuel:1993it,Henderson:1996ez,Adhikari:1997dz,Henderson:1997kv,PHILLIPS1998255,Cavalcanti:1998jx,Mitra:1998vr,Ferkous:2013uqa,ANWONG20182547,Altunkaynak:2006ik,Erman:2010sk,Akbas:2023ucf,Erman:2024wal}), we develop a systematic framework to derive the energy–eigenvalue equation for characteristic topologies, drawing on the classification of flat-space universe models \citep{Riazuelo:2003ud,COMPACT:2022gbl} and  adapting lattice-field-theory techniques—already used to study finite-volume effects (see, e.g., \citep{Antonelli:2000mc,Bernard:2008zza,Bernard:2010fp,Hammer:2017uqm,Romero-Lopez:2018zyy,Romero-Lopez:2020rdq,Muller:2020vtt,Garofalo:2021bzl,Bubna:2023oxo,Meissner:2014dea})  on energy eigenvalues\citep{Antonelli:2000mc} in spaces such as the torus—to handle mode sums\citep{Romero-Lopez:2018zyy}. Finally, in Section~\ref{sectioniv} we demonstrate applications of these results at cosmological scales.


 \section{Mathematical Prerequisites}\label{sectionii}

  At a given cosmic time, consider a spacelike three-dimensional hypersurface $\Sigma$ with the spatial geometry induced by the spacetime metric in Eq.~(\ref{frw}). Each such geometry describe the universal covering space~\citep{Lachieze-Rey:1995qrb,Frankel:1997ec,Levin:2001fg}.  For a detailed review and mathematical background on the topology of the universe, see~\citet{Lachieze-Rey:1995qrb}.

  \subsection{Universal Covering Space}
  Homotopy\citep{Frankel:1997ec} defines an equivalence relation on the set of curves from \( x \) to \( y \), where two curves are considered equivalent if one can be continuously deformed into the other while keeping the endpoints fixed\footnote{Let \( \Sigma \) be a connected manifold, and let \( x, y \in \Sigma \). Consider two continuous curves \( \gamma, \gamma': [0,1] \to \Sigma \) with \( \gamma(0) = \gamma'(0) = x \) and \( \gamma(1) = \gamma'(1) = y \). We say that \( \gamma \) and \( \gamma' \) are homotopic if there exists a continuous function \( F: [0,1] \times [0,1] \to \Sigma \) such that \( F(0, t) = \gamma(t) \), \( F(1, t) = \gamma'(t) \) for all \( t \in [0,1] \), and \( F(s,0) = x \), \( F(s,1) = y \) for all \( s \in [0,1] \). \( \Sigma \) is simply connected if, for any \( x, y \in \Sigma \), all curves connecting \( x \) and \( y \) are homotopic.  }. 
  
  Choose a base point \( x \) in \( \Sigma \) and consider the paths from \( x \) to  \( x \) (loops)\footnote{The set of loops through $x$ can be given a natural group structure i.e., the fundamental group $\pi_1(\Sigma)$\citep{Wald:1984rg,Frankel:1997ec,Lachieze-Rey:1995qrb}. }. Consider any path from the base point \( x \) to a point \( y \in \Sigma \).  Each such path belongs to a homotopy class defined by loops based at \( x \).
  If the space is simply connected, the universal covering space $\tilde{\Sigma}$ coincides with the original space \( \Sigma \) (there is a natural map \( \hat{f} : \tilde{\Sigma} \to \Sigma \), which is one-to-one if and only if \( \Sigma \) is simply connected, and this map takes each \( \tilde{y} \in \tilde{\Sigma} \) to the endpoint \( y \in \Sigma \) of a curve from \( x \) to \( y \), where the curve belongs to the specific equivalence class of curves associated with \( \tilde{y} \)\citep{Wald:1984rg}), as there is only one homotopy class of loops based at \( x \), and thus all curves from \( x \) to \( y \) are homotopic. However, if the space is multiply connected, for the same fixed points \( x \) and \( y \), there will be paths belonging to different homotopy classes, and the same \( y \) will correspond to multiple points \( \tilde{y}, \tilde{y}', \dots \) in $\tilde{\Sigma}$.

  The \textit{universal covering space} \( \tilde{\Sigma} \) is constructed as follows\citep{Wald:1984rg,Frankel:1997ec,Lachieze-Rey:1995qrb}:  For a fixed base point \( x \), each \( \tilde{y} \in \tilde{\Sigma} \) corresponds to the endpoint of a curve from \( x \) to \( y \in \Sigma \), where the curve belongs to an equivalence class of loops based at \( x \). 
 The entire space \( \tilde{\Sigma} \) is defined from all those $\tilde{y}$ i.e., the set of all equivalence classes of paths from a base point \( x \) to points \( y \in \Sigma \), as \( y \) ranges over the entire $\Sigma$.

Consider a point \( x \in \Sigma \) and a loop \( \gamma \) based at \( x \). If \( \gamma \) lies entirely within a simply connected domain of \( \Sigma \), then the pair \( (x, \gamma) \) generates a single point \( \tilde{x} \) in the universal covering space \( \tilde{\Sigma} \). Otherwise, \( (x, \gamma) \) generates additional points \( \tilde{x}', \tilde{x}'', \dots \) in \( \tilde{\Sigma} \), which are said to be \emph{homologous} to \( \tilde{x} \).
The transformations \( \tilde{x} \to \tilde{x}' \), \( \tilde{x} \to \tilde{x}'' \), etc., form a group of isometries\footnote{  For a manifold \( \Sigma \) endowed with a metric \( g \), a group of transformations of \( \Sigma \) is a group of isometries if each transformation in the group preserves the metric \( g \).  } which is called the \emph{holonomy group} \( \Gamma \) of \( \tilde{\Sigma} \) (see also \citep{Lachieze-Rey:1995qrb}). We can write 
$\Sigma = \tilde{\Sigma} / \Gamma$, where $\tilde{\Sigma}$ is the simply connected 
universal covering space, and \( \Gamma \) is also a discrete group of isometries of \( \tilde{\Sigma} \) such that no non-identity element of \( \Gamma \) has a fixed point in \( \tilde{\Sigma} \) (i.e., each isometry moves every point of \( \tilde{\Sigma} \) without leaving any point unchanged). 

Given that \( \tilde{\Sigma} \) is endowed with the Riemannian metric \( \tilde{g} \) and that \( \Gamma \) is realized as a group of isometries preserving \( \tilde{g} \), if \( \Gamma \) acts on \( \tilde{\Sigma} \) in a fixed-point free and properly discontinuous manner\footnote{A group \( \Gamma \) is \emph{fixed-point free} if every non-identity isometry has no fixed points. It is \emph{properly discontinuous} if each point \( x \in \tilde{\Sigma} \) has a neighborhood whose images under all isometries in \( \Gamma \) are pairwise disjoint.}, then  the  space \(\Sigma= \tilde{\Sigma}/\Gamma \) inherits a well-defined metric induced by \( \tilde{g} \).

 \subsection{Flat Cosmic Topologies}

When $\tilde{\Sigma} = \mathbb{R}^3$ has its usual (flat) metric
\[
R^2d\sigma^2 \equiv R^2 \bigl[d\chi^2 + \chi^2(d\theta^2 + \sin^2\theta\,d\phi^2)\bigr],
\]
its full isometry group is $ \mathbb{R}^3 \times \mathrm{SO}(3)$. 
Discrete subgroups \(\Gamma\) of \(\mathbb{R}^3 \times \mathrm{SO}(3)\) (without fixed points) generate 18 possible locally Euclidean spaces, 17 of which are multiply connected\citep{Lachieze-Rey:1995qrb}. These spaces are constructed via transformations such as parallel translations, rotations about arbitrary axes followed by parallel translations (corkscrew motions), reflections across planes with parallel translations (glide reflections), and specific combinations of these transformations.

\subsection{$\Gamma$-Periodic Wavefunctions on \(L^2\left(\mathbb{R}^3\right)\)}

Any \(\Gamma\)-periodic function in \(L^2(\tilde{\Sigma})\) (the space of square-integrable functions defined on the simply-connected space \(\tilde{\Sigma}\)) can be identified with a function in \(L^2(\tilde{\Sigma}/\Gamma)\). Since \(\tilde{\Sigma}\) is simply connected, a wavefunction defined on \(\tilde{\Sigma}\) would be single-valued. However, it would correspond to a multivalued wavefunction on \(\tilde{\Sigma}/\Gamma\) for various elements of \(\Gamma\). For physical reasons, we will ensure that the wavefunction on \(\Sigma=\tilde{\Sigma}/\Gamma\) is single-valued by choosing a specific branch of the $\Gamma-$periodic wavefuntion on $\tilde{\Sigma}=\mathbb{R}^3$(see also \citet{Dowker:1972np}).

 Consider a free particle of mass $m$ confined to the multiply connected space $\Sigma = \mathbb{R}^3/\Gamma$, whose stationary wavefunctions satisfy the time independent  Schrödinger  equation $\nabla^2\Psi(\mathbf{x}) + k^2\Psi(\mathbf{x}) = 0$.  By lifting the Schrödinger equation on $\Sigma$ to its universal cover $\tilde\Sigma = \mathbb{R}^3$, one sees that each eigenmode on $\Sigma$ corresponds uniquely to a $\Gamma$-invariant eigenmode on $\mathbb{R}^3$, so that finding the spectrum reduces to selecting those solutions on $\mathbb{R}^3$ which remain invariant under the holonomy group $\Gamma$; this construction, first presented in \cite{Riazuelo:2003ud} and later extended to arbitrary geometric parametrizations and observer locations in \cite{COMPACT:2023rkp}, ensures that both the eigenfunctions and their eigenvalues faithfully encode the topology of $\Sigma$.


\section{A Toy model: A Simple Quantum System in Flat Cosmic-Inspired  Topologies}\label{sectioniii}

Here, we will introduce a toy model appropriate to study the impact of the topology of space on the energy eigenvalues.
In three dimensions, the Dirac delta potential is given by \( V(\mathbf{x}) = -\epsilon \delta^{(3)}(\mathbf{x}) \) ($\epsilon>0$), where we define the parameter \( g \equiv \frac{m\epsilon}{\hbar^2} \), which has units of  length \([L]\). The associated Schrödinger equation is given by

\begin{equation}\label{schrondig}
    \nabla^2 \Psi(\mathbf{x})+2g \delta^{(3)}(\mathbf{x}) \Psi(\mathbf{x}) = -2\tilde{E} \Psi(\mathbf{x}).
\end{equation}
where \( \tilde{E} \equiv \frac{m E}{\hbar^2} \) has units of inverse length squared \([L]^{-2}\).

\subsection{ Dirac Delta potential in $\mathbb{R}^3$}

The wavefunctions defined over the entire space $\mathbb{R}^3$ can be expressed as a continuous superposition of plane waves through the Fourier transform:
\begin{equation}\label{transform1}
    \tilde{\Psi}(\mathbf{k}) = \frac{1}{(2\pi)^{3/2}} \int_{\mathbb{R}^3} e^{-i\mathbf{k} \cdot \mathbf{x}} \Psi(\mathbf{x}) \, d^3 x.
\end{equation}
The original wavefunction is recovered via the inverse Fourier transform:
\begin{equation}\label{fourier2}
    \Psi(\mathbf{x}) = \frac{1}{(2\pi)^{3/2}} \int_{\mathbb{R}^3} e^{i\mathbf{k} \cdot \mathbf{x}} \tilde{\Psi}(\mathbf{k}) \, d^3 k.
\end{equation}

Applying Eqs.~(\ref{transform1}) and~(\ref{fourier2}) to the term in Eq.~(\ref{schrondig}) that involves the Dirac delta function, we find:
\begin{equation}\label{ytox}
    \delta^{(3)}(\mathbf{x}) \Psi(\mathbf{x}) = \frac{1}{(2\pi)^{3/2}} \int d^3k' \, e^{-i\mathbf{k'}\cdot \mathbf{x}} \left[\frac{\Psi(\mathbf{0})}{(2\pi)^{3/2}} \right].
\end{equation}

Now, using Eqs.~(\ref{fourier2}) and~(\ref{ytox}), we can express all terms in Eq.~\eqref{schrondig} in momentum space, yielding:
\begin{equation}\label{fouriertransformed}
    \int d^3k' \left[ \left(\frac{k'^2}{2} - \tilde{E}\right)\tilde{\Psi}(\mathbf{k'}) - \frac{g}{(2\pi)^{3/2}}\Psi(\mathbf{0}) \right] e^{i\mathbf{k'} \cdot \mathbf{x}} = 0.
\end{equation}
From Eq.~(\ref{fouriertransformed}), we deduce that 
\begin{equation}
    \tilde{\Psi}(\mathbf{k}) = \frac{g}{(2\pi)^{3/2}} \frac{\Psi(\mathbf{0})}{\frac{k^2}{2} - \tilde{E}}.
\end{equation}
We determine the energy eigenvalue for the bound state by setting \( \tilde{E} = -|\tilde{E}| \). Evaluating Eq.~(\ref{fourier2}) at \( \mathbf{x} = \mathbf{0} \) and assuming \( \Psi(\mathbf{0}) \neq 0 \), we obtain
\begin{equation}\label{renorm}
   \pi^2 = g \int_{0}^{\infty} dk' \frac{k'^2}{k'^2 +2 |\tilde{E}|},
\end{equation}
which the right hand side diverges.  The bound state's existence is ensured by requiring the above relation (Eq.(\ref{renorm})) to hold simultaneously for finite cutoff $\Lambda \equiv k_{\mathrm{max}}$ and in the $\Lambda \to \infty$ limit. This implies that $ \pi^2 =g(\Lambda) \int_{0}^{\Lambda} dk' \frac{k'^2}{k'^2 +2 |\tilde{E}|}$ provides a regularization mechanism, as it leads to \( g \equiv g(\Lambda) \), i.e.,
\begin{equation}\label{physicalcondition}
   \frac{1}{g(\Lambda)} = \frac{\Lambda}{\pi^2}-\frac{\sqrt{2|\tilde{E}|}}{\pi^2}\arctan{\left(\Lambda/\sqrt{2|\tilde{E}|}\right)}.
\end{equation}

From Eq.~(\ref{physicalcondition}), taking the limit \( \Lambda \to \infty \) leads to \( g \to 0 \), rendering Eq.~(\ref{renorm}) indeterminate. To obtain a well-defined renormalized coupling, we redefine the parameter as \( g \to 2\pi g_R \), where \( g_R \) is determined by the following condition:
\begin{equation}
    \frac{1}{g_R} =2\pi \lim_{\Lambda\to \infty} \left(\frac{\Lambda}{\pi^2} - \frac{1}{g(\Lambda)}\right) = \sqrt{2|\tilde{E}|},
\end{equation}
which leads to\citep{Cavalcanti:1998jx}
\begin{equation}
    \tilde{E} =-|\tilde{E}|= -\frac{1}{2g^2_{R}}\implies E=-\frac{\hbar^2}{2 m g_R^2}
\end{equation}

Note that we are not concerned with the specific energy eigenvalue itself, but rather with how it varies across different topologies.

\subsection{Dirac Delta potential in Flat Cosmic-Inspired  Topologies $E_1-E_2$}

 We introduce a toy model consisting of a three-dimensional Dirac delta potential and calculate the energy eigenvalues of a bound state within cosmic-inspired topologies. Although this toy model lacks physical realism, it offers valuable intuition and showcases our analytical approach for each topological case. We then investigate how these topological effects modify the energy levels of the quantum particle in its renormalized bound state.
 The  renormalization procedure of the bound state problem involving Dirac-delta interactions on Riemannian manifolds has been extensively studied in \citet{Altunkaynak:2006ik}.

For our purposes, it suffices to solve Eq.(\ref{schrondig}) specifically for two compact and orientable topologies, which can be visualized by identifying corresponding faces of a parallelepiped \citep{Riazuelo:2003ud,Fujii:2011ga,COMPACT:2022gbl}(each is defined by two to six parameters governing face shapes, translations, and rotations, and three to six parameters may be required to specify an observer’s position and orientation; see \citep{COMPACT:2023rkp}):

\begin{itemize}
\item \textbf{3-Torus} $(E_1):$ Opposite faces are identified by translations.

    \item \textbf{Half-Turn space} $(E_2):$  The  faces along the $z-$direction are identified through a combination of a $\pi$ rotation about the $z$-axis and a translation  along the $z$-axis. The  faces in the $x$-direction and the  faces in the $y$-direction are identified through pure translations.


\end{itemize}

\subsubsection{Systematic Approach to Derive the Equation for the Energy Eigenvalue }

The eigenmodes of the Laplacian and the classification of universe topologies corresponding to flat geometries are discussed in~\cite{Riazuelo:2003ud,COMPACT:2022gbl}, while lattice field theory techniques are applied to perform the necessary summations~\cite{Romero-Lopez:2018zyy}. Building upon these results, we develop a systematic method to derive the energy eigenvalue equation for characteristic topologies starting from Eq.~(\ref{schrondig}). We focus on compact flat topologies ($E_1$--$E_2$) and compute explicit spectral shifts using a Dirac delta potential as a simplified theoretical model.

Consider a particle confined within a three-dimensional space of non-trivial topology, subject to the potential $V(\mathbf{x}) = -\epsilon\delta^{(3)}(\mathbf{x})$.
We assume the fundamental domain to be a cubic region of side physical length \( L \), i.e., \( L_1 = L_2 = L_3 = L \).  Assume that the Fourier basis compatible with the space's topology is denoted by $\xi_{\mathbf{k}}(\mathbf{x})$. 
Then the Fourier series expansion of a function $f(\mathbf{x})$ and its coefficients take the form:
\begin{align}
    f(\mathbf{x}) &=  \sum_{\mathbf{k}} \tilde{f}_{\mathbf{k}} \,\xi_{\mathbf{k}}(\mathbf{x}), \label{eq:fourier_expansion1} \\
    \tilde{f}_{\mathbf{k}} &=  \int_{[-\frac{L}{2},\frac{L}{2}]^3} f(\mathbf{x}) \xi_{\mathbf{k}}^*(\mathbf{x}) \, d^3x, \label{fourrierhalf2}
\end{align}
These basis functions satisfy the orthonormality relation:
\begin{equation}
    \int_{[-\frac{L}{2},\frac{L}{2}]^3} \xi_{\mathbf{k}}(\mathbf{x}) \xi_{\mathbf{k}'}^*(\mathbf{x}) \, d^3x = \delta_{\mathbf{k},\mathbf{k}'}
\end{equation}

We apply the discrete Fourier transform, as defined in Eqs.~(\ref{eq:fourier_expansion1}) and~(\ref{fourrierhalf2}), to the Schrödinger equation (Eq.~\ref{schrondig}), obtaining the following expression (see also~\cite{Altunkaynak:2006ik}):

\begin{equation}
    \sum_{\mathbf{k}}\xi_{\mathbf{k}}\left[-\mathbf{k}^2\tilde{\Psi}_{\mathbf{k}}+2g\Psi(\mathbf{0})\xi_{\mathbf{k}}^*(\mathbf{0})+2\tilde{E}\tilde{\Psi}_{\mathbf{k}}\right]=0,
\end{equation}
which will give
\begin{equation}
    \tilde{\Psi}_{\mathbf{k}}=2g\frac{\Psi(\mathbf{0})}{\mathbf{k}^2+2|\tilde{E}|}\xi_{\mathbf{k}}^*(\mathbf{0})
\end{equation}
 By using that $\Psi(\mathbf{0})=\sum_{\mathbf{k}}\tilde{\Psi}_{\mathbf{k}}\xi_{\mathbf{k}}(\mathbf{0})$ and that $\Psi(\mathbf{0})\neq\mathbf{0}$,   we get
 
\begin{equation}\label{condition3dtor}
    \sum_{\mathbf{k}}\frac{|\xi_{\mathbf{k}}(\mathbf{0})|^2}{\mathbf{k}^2+2|\tilde{E}|}=\frac{1}{2g}.
\end{equation}

\subsubsection{Validation of Our Approach: Dirac delta potential on the circle}

In one dimension, the time-independent Schrödinger equation for a particle of mass $m$ is given by  
\begin{equation}\label{schrondiger1d}
    -\frac{\hbar^2}{2m} \frac{\rm d^2}{\rm dx^2} \Psi(x) + V(x) \Psi(x) = E \Psi(x).
\end{equation}
We consider the potential $V(x) = -\epsilon\delta(x)$, define the positive parameter $g \equiv \frac{m\epsilon}{\hbar^2}$, and introduce the characteristic length scale $\alpha \equiv 1/g$.
We analyze the existence of a single bound state and express the problem in terms of the parameter $\tilde{E} \equiv \frac{m E}{\hbar^2}$, which has the scale of inverse length squared.

For the case where \( x \in \left[-\frac{L}{2}, \frac{L}{2}\right] \) and \( x = R \theta \), with \( R = \frac{L}{2 \pi} \), and considering periodic boundary conditions \( \Psi\left(-\frac{L}{2}\right) = \Psi\left(\frac{L}{2}\right) \), then the  parameter \( \tilde{E} \)  for the bound state ($\tilde{E}=-|\tilde{E}|$) satisfies the equation:
\begin{equation}\label{discreteenergy}
   \sqrt{2 |\tilde{E}|}= g \coth{\left( \frac{ \sqrt{|\tilde{E}|}L}{\sqrt{2}} \right)} .
\end{equation}
Any periodic  square-integrable function $\Psi$ of the Hilbert space (with dot product $<\Psi|\Phi>=\int_{-L/2}^{L/2}\Psi^*(x)\Phi(x) dx$) can be expanded in a Fourier series using the complete orthonormal basis functions
\begin{equation}\label{eigenfunctions}
    \xi_n(x) = \frac{1}{\sqrt{L}} e^{\frac{2\pi i n x}{L}}, \quad n \in \mathbb{Z}.
    \end{equation}
Note that Eq.~(\ref{eigenfunctions}) forms an orthonormal basis as the eigenfunctions of the self-adjoint (for a detailed discussion on self-adjointness, see \citep{Hall:2013jtz}) operator \(  k = \frac{p}{\hbar}=-i \frac{d}{dx} \), with eigenvalues given by \( k = \frac{2\pi }{L}n \).

Eq. \eqref{discreteenergy} can be derived by taking the discrete Fourier transform of equation \eqref{schrondiger1d}, leading to:
\begin{equation}\label{psintilde}
    \tilde{\Psi}_n=\frac{2g}{L^{1/2}}\frac{\Psi(0)}{k^2+2|\tilde{E}|}.
\end{equation}
Given that 
\begin{equation}
    \Psi(0)=\frac{1}{L^{1/2}}\sum_{n=-\infty}^{\infty}\tilde{\Psi}_n,
\end{equation}
substituting Eq.~\eqref{psintilde} into the above expression yields:
\begin{equation}
    \frac{L}{2g}=\sum_{n=-\infty}^{\infty}\frac{1}{k_n^2+2|\tilde{E}|}=\frac{L}{2\sqrt{2|\tilde{E}|}}\coth{\left( \frac{ |\tilde{E}|^{\frac{1}{2}}L}{2^{\frac{1}{2}}} \right)},
\end{equation}
from which equation \eqref{discreteenergy} follows (see also Appendix \ref{App}). To estimate the magnitude of corrections to a bound energy eigenvalue in the large-L limit ($\alpha\ll L$), we approximate using Eq.~\eqref{discreteenergy}:
\begin{equation}\label{approximation1d}
   |\tilde{E}|\simeq \frac{1}{2\alpha^2}\left(1 + 4e^{-L/\alpha}\right).
\end{equation}


\subsubsection{Renormalized Bound State for a 3D Dirac Delta Potential in 3-Torus ($E_1$)}

A simple example is the flat 3-Torus.
For example, the fundamental domain corresponds to the cubic region 
 $ \{(x,y,z) \in \mathbb{R}^3 \mid -L/2 \leq x,y,z \leq L/2\}$
with the observer positioned at the cube's center. The repeated copies of this cube tile $\mathbb{R}^3$, inducing the torus's flat geometry.

Assume a particle confined in a dirac delta potential $V(\mathbf{x})=-\epsilon\delta^{(3)}(\mathbf{x})$ within  the three-dimensional flat torus. 
In the 3-torus topology, the Fourier basis is given by~\citep{Riazuelo:2003ud}
\begin{equation}\label{basistorus}
    \xi_{\mathbf{k}}(\mathbf{x}) = \frac{1}{L^{3/2}}\,e^{i\mathbf{k}\cdot \mathbf{x}},
    \quad 
    k_i = \frac{2\pi n_i}{L}, 
    \; n_i \in \mathbb{Z}.
\end{equation}

Equation~\eqref{condition3dtor} can then be expressed as
\begin{equation}\label{lhs}
   \sum_{\mathbf{n}}\frac{1}{\mathbf{n}^2+l}=\frac{2\pi^2 L}{g}
\end{equation}
where $l\equiv L^2|\tilde{E}|/2\pi^2$ and $\mathbf{n}=(n_x,n_y,n_z)$.

The LHS of Eq.(\ref{lhs}) yields (notably, identical sums appear in lattice field theory; see \cite{Romero-Lopez:2020rdq,Meißner_Rusetsky_2022} for instance):
\[
\sum_{\mathbf{n}}\frac{1}{\mathbf{n}^2+l}=\sum_{\mathbf{n}}  \int d^3\tilde{x}  \frac{\delta^{(3)}(\tilde{\mathbf{x}} - \mathbf{n})}{\tilde{x}^2 + l}=
\]

\[
 = \int d^3\tilde{x} \frac{1}{\tilde{x}^2 + l} + \sum_{\mathbf{n} \neq 0} \int d^3\tilde{x}\,  \frac{e^{i2\pi \mathbf{n} \cdot \tilde{\mathbf{x}}}}{\tilde{x}^2 + l},
\]

We have used the Poisson summation formula (see Lemma~I in Appendix~\ref{AppB})
\begin{equation}\label{identitytorus}\frac{1}{L^3}\sum_{\mathbf{n}}e^{\frac{2\pi i}{L}\mathbf{n}\cdot\mathbf{x}}=\sum_{\mathbf{n}}\delta(\mathbf{x}-\mathbf{n}L).\end{equation}
Furthermore, note that 
\[
\sum_{\mathbf{n} \neq 0} \int d^3\tilde{x}\,  \frac{e^{i2\pi \mathbf{n} \cdot \tilde{\mathbf{x}}}}{\tilde{x}^2 + l} =\pi\sum_{\mathbf{n} \neq 0} \frac{e^{-2\pi n \sqrt{l}}}{n },
\]
where we denote \( n \equiv |\mathbf{n}| \) (and analogously for \( \mathbf{x} \)). Additionally, we define the integral
\[
I(\lambda) = 4\pi \int_0^\lambda d\tilde{x} \, \frac{\tilde{x}^2}{\tilde{x}^2 + l}=4\pi\left[\lambda-\sqrt{l}\arctan\left(\lambda/\sqrt{l}\right)\right],
\]
which arises from the expression \( \int d^3\tilde{x} \, \frac{1}{\tilde{x}^2 + l} \). Note that \( I(\lambda) \) diverges as \( \lambda \to \infty \). 
 The term \(\sim \lambda\) can be absorbed into the running coupling constant \(g_R\), as done previously. 
Specifically, we set the renormalized coupling constant as 
\begin{equation*}
     \frac{1}{g_R} = 2\pi\lim_{\lambda\to \infty} \left(\frac{2\lambda}{\pi L}-\frac{1}{g(\lambda)} \right)  
\end{equation*}
Hence, one finds
    \begin{equation}\label{enegryeigenstatestorus}
    \frac{1}{g_R}  = \sqrt{2|\tilde{E}|} -\frac{1}{L}\sum_{\mathbf{n} \neq 0} \frac{e^{-n \sqrt{2|\tilde{E}|}L}}{n }.
\end{equation}
In the large-\(L\) limit, we assume that only the term with \(n = 1\) contributes to the sum at leading order. Given that the possible equivalent cases are \(\mathbf{n} = (\pm1,0,0), (0,\pm1,0), (0,0,\pm1)\), Eq.~(\ref{enegryeigenstatestorus}) then admits the following approximation:
\begin{equation}\label{approximation3Dtori}
    |\tilde{E}| \simeq \frac{1}{2g_R^2} + \frac{6}{g_RL} e^{-L/g_R}.
\end{equation}

\subsubsection{Renormalized Bound State for a 3D Dirac Delta Potential in Half-turn space ($E_2$)}
The half-turn space is formed by taking a cube and identifying one pair of opposite faces (aligned with the $z$-axis) with a $\pi$-rotation, while the remaining pairs of opposite faces are identified directly, analogous to the construction of a 3-torus. The eigenfunctions are obtained as linear combinations of plane-wave solutions on the universal cover $\mathbb{R}^3$, carefully combined to ensure invariance under the corresponding group. 

Such combinations explicitly satisfy the necessary periodic boundary conditions arising from the underlying topological identifications.  In such a case, the symmetries for wavefunctions are:
\begin{align}
    \Psi(x-L/2,y,z) &= \Psi(x+L/2,y,z), \label{eq:bc_x} \\
    \Psi(x,y-L/2,z) &= \Psi(x,y+L/2,z), \label{eq:bc_y} \\
    \Psi(x,y,z-L/2) &= \Psi(-x,-y,z), \label{eq:bc_z}\\
    \Psi(x,y,z-L/2) &= \Psi(x,y,z+L/2).
\end{align}

Furthermore, let us define the  sets
\[
\mathcal{I}_0 = \left\{ \mathbf{n} \in \mathbb{Z}^3 \,\middle|\, \mathbf{n} = (0, 0, n_z),\; n_z \in 2\mathbb{Z} \right\},
\]
\[
 \mathcal{I}^*= \left\{ \mathbf{n} \in \mathbb{Z}^3 \,\middle|\, 
\begin{array}{l}
\, n_x \in \mathbb{Z}^+,\, n_y \in \mathbb{Z},\,n_z \in 2\mathbb{Z} \\
\text{or } n_x = 0,\, n_y \in \mathbb{Z}^+,\, n_z \in 2\mathbb{Z}
\end{array}
\right\},
\]
\[
 \mathcal{I}= \left\{ \mathbf{n} \in \mathbb{Z}^3 \,\middle|\, 
\begin{array}{l}
\, n_x \in \mathbb{Z}^+,\, n_y,n_z \in \mathbb{Z} \\
\text{or } n_x = 0,\, n_y \in \mathbb{Z}^+,\, n_z \in \mathbb{Z}
\end{array}
\right\},
\]

The eigenmodes of the Laplacian operator $\nabla^2$, form an orthonormal basis and satisfying $\nabla^2\xi_{\mathbf{k}} = -\mathbf{k}^2\xi_{\mathbf{k}}$\citep{Riazuelo:2003ud}:
\begin{widetext}
    \begin{equation}\label{basishalf}
    \xi_{\mathbf{k}}(\mathbf{x})=\frac{e^{ik_z z}}{L^{3/2}} 
   \times\begin{cases} \frac{1}{\sqrt{2}} \left[ e^{i(k_x x + k_y y)} + (-1)^{n_z} e^{-i(k_x x + k_y y)} \right], & \, \text{when } \mathbf{n}\in\mathcal{I} \\
    1, & \ \text{when } \mathbf{n}\in\mathcal{I}_0
   \end{cases}
\end{equation}
\end{widetext}
with wavevectors given by $k_i = \frac{2\pi n_i}{L}$.

 Next, we apply the Fourier transform to the Schrödinger equation (Eq.~\ref{schrondig}) and derive   the  equation Eq.(\ref{condition3dtor}) with $k_i$ defined as previously. We get:
\begin{equation}\label{conditionhalfturn}
  \frac{1}{2} \sum_{\mathbf{n}\in \mathcal{I}}\frac{\left[1+(-1)^{n_z}\right]^2}{\mathbf{n}^2+l}=\frac{2\pi^2 L}{ g} 
\end{equation}
where $\mathbf{n}=(n_x,n_y,n_z)$ where $  n_x \in \mathbb{Z}^+, n_y, n_z \in \mathbb{Z} \text{ or } n_x = 0, n_y \in \mathbb{Z}^+, n_z \in \mathbb{Z}$.  The summation in Eq.~\eqref{conditionhalfturn}  yields
\begin{equation}\label{eqsumhalfnnot0}
   \sum_{\mathbf{n}\in \mathcal{I}^*}\frac{1}{\mathbf{n}^2+l}=\frac{\pi^2 L}{ g} .
\end{equation}

Given that the complete sum is obtained by evaluating two distinct contributions:  
(i) modes satisfying either \( n_x \in \mathbb{Z}^+,\, n_y \in \mathbb{Z} \) or \( n_x = 0,\, n_y \in \mathbb{Z}^+ \), with \( n_z \in 2\mathbb{Z} \) in both cases; and  
(ii) the  case \( n_x = n_y = 0 \) with \( n_z \in 2\mathbb{Z} \),  
we now repeat the Fourier transform procedure for the remaining \(\mathbf{n}\)-modes, restricting to the case \( n_x = n_y = 0 \) and \( n_z \in 2\mathbb{Z} \) (see Eq.~\eqref{basishalf}), which yields
\begin{equation}\label{zeromodes}
    \sum_{\mathbf{n}\in\mathcal{I}_0}\frac{1}{\mathbf{n}^2+l}=\frac{2\pi^2L}{g}
\end{equation}

The total sum over all modes is then obtained by adding Eq.~\eqref{eqsumhalfnnot0} and Eq.~\eqref{zeromodes}:

\begin{equation}\label{halfturnconsition}
    \sum_{\mathbf{n}\in\mathcal{I}_0}\frac{1}{\mathbf{n}^2+l}+\sum_{\mathbf{n}\in\mathcal{I}^*}\frac{1}{\mathbf{n}^2+l}=\frac{3\pi^2L}{g}
\end{equation}
Now, given that
\[
\sum_{\mathbf{n} \in \mathcal{I}_0} \frac{1}{\mathbf{n}^2 + l}
+ \sum_{\mathbf{n} \in \mathcal{I}^*} \frac{1}{\mathbf{n}^2 + l}
= \sum_{\mathbf{n} \in \mathcal{I}_0 \cup \mathcal{I}^*} \frac{1}{\mathbf{n}^2 + l},
\]
and if we use  the identity (see Lemma~II in Appendix~\ref{AppB}):
\begin{widetext}
\begin{equation} \label{diracdeltahalf}
    \sum_{\mathbf{n} \in \mathcal{I}_0\cup\mathcal{I}^*} \delta^{(3)}(\mathbf{x} - \mathbf{n}L)
    = \frac{1}{L^3} \left[
        \sum_{\mathbf{n} \in \mathcal{I}^*} \left(
            e^{i \frac{2\pi}{L} \mathbf{n}_1 \cdot \mathbf{x}} 
            + e^{i \frac{2\pi}{L} \mathbf{n}_2 \cdot \mathbf{x}}
        \right) 
        + \sum_{\mathbf{n} \in \mathcal{I}_0} e^{i \frac{2\pi}{L} \mathbf{n} \cdot \mathbf{x}}
    \right],
\end{equation}
\end{widetext}
we obtain the following expression from the left-hand side of Eq.~\eqref{halfturnconsition}:
\[
 \sum_{\mathbf{n}\in\mathcal{I}_0\cup\mathcal{I}^*}\frac{1}{\mathbf{n}^2+l}=\int d^3\tilde{x}  \frac{\sum_{\mathbf{n}\in\mathcal{I}_0\cup\mathcal{I}^*} \delta^{(3)}(\tilde{\mathbf{x}} - \mathbf{n})}{\tilde{x}^2 + l}=
\]
\begin{equation}\label{integrhalf}
    =\int \frac{d^3\tilde{x} }{\tilde{x}^2 + l}+\sum_{ \mathbf{n}\in \mathcal{I}_0\setminus \{\mathbf{0}\}}\int d^3\tilde{x}\, \frac{e^{i 2\pi \boldsymbol{n} \cdot \tilde{\mathbf{x}} }}{\tilde{x}^2 + l} + \sum_{\mathbf{n} \in\mathcal{I}^*} \int d^3\tilde{x}\, \frac{C_{\mathbf{n}}}{\tilde{x}^2 + l}.
\end{equation}
We have defined
\begin{equation}\label{rewritedelta}
    C_{\mathbf{n}} \equiv e^{i 2\pi \boldsymbol{n}_1 \cdot \tilde{\mathbf{x}}} + e^{i 2\pi \boldsymbol{n}_2 \cdot \tilde{\mathbf{x}}},
\end{equation}
where   \(\boldsymbol{n}_1 = (n_x, n_y, n_z)\) and \(\boldsymbol{n}_2 = (-n_x, -n_y, n_z)\).


The first integral in Eq.~\eqref{integrhalf} is evaluated through renormalization, following a similar procedure as before.
The second and third integrals in Eq.~\eqref{integrhalf} are computed as follows (see Eq.~\eqref{rewritedelta}):
\begin{align}
    \int d^3\tilde{x}\,\frac{C_{\mathbf{n}}}{\tilde{x}^2 + l} 
    &= 2\pi \frac{e^{-2\pi n \sqrt{l}}}{n}, \\
    \int d^3\tilde{x}\,\frac{e^{i2\pi n_z z}}{\tilde{x}^2 + l} 
    &= \pi \frac{e^{-2\pi |n_z| \sqrt{l}}}{|n_z|},
\end{align}
where \( n  = \sqrt{n_x^2 + n_y^2 + n_z^2} \).

Finally, we set a renormalized coupling constant through the following condition
\begin{widetext}
    \begin{equation}\label{1overgrhts}
    \frac{1}{g_R} = 3\pi\lim_{\lambda\to \infty} \left(\frac{4\lambda}{3\pi L}-\frac{1}{g(\lambda)} \right) =\sqrt{2|\tilde{E}|}-\frac{1}{L}  \sum_{n_z\in 2\mathbb{Z}\setminus \{0\}}\frac{e^{-|n_z| \sqrt{2|\tilde{E}|}L}}{|n_z|}-\frac{2}{L}  \sum_{\mathbf{n} \in \mathcal{I}^*}\frac{e^{-n \sqrt{2|\tilde{E}|}L}}{n}.
\end{equation}
\end{widetext}
The second term corresponds to modes with \( n_x = n_y = 0 \) and \( n_z \in 2\mathbb{Z} \setminus \{0\} \), while the last term is constrained by \( n_x \in \mathbb{Z}^+ \), \( n_y \in \mathbb{Z} \) or \( n_x = 0 \), \( n_y \in \mathbb{Z}^+ \), with \( n_z \in 2\mathbb{Z} \) in all cases.
The following sum has a closed form:
\begin{equation}\label{closedformsum}
    \sum_{n_z\in 2\mathbb{Z}\setminus \{0\}}\frac{e^{-|n_z| \sqrt{2|\tilde{E}|}L}}{|n_z|}=-\ln\left(1-e^{-2\sqrt{2|\tilde{E}|}L}\right),
\end{equation}
and thus Eq.~(\ref{1overgrhts}) (given Eq.(\ref{closedformsum})) becomes
\begin{equation}\label{eqconditionhalfturngr}
     \frac{1}{g_R} = \sqrt{2|\tilde{E}|}+  \frac{1}{L}\left[ \ln(1-e^{-2\sqrt{2|\tilde{E}|}L})-  2\sum_{\mathbf{n} \in \mathcal{I}^*}\frac{e^{-n \sqrt{2|\tilde{E}|}L}}{n}\right].
\end{equation}

 In the large-\(L\) limit, and to leading order, the sum receives
contributions solely from the \(n = 1\) term.
Consequently, Eq.~\eqref{eqconditionhalfturngr} can be approximated as
\begin{equation}\label{approximation3Dhalfturn}
    |\tilde{E}| \simeq \frac{1}{2g_R^2} + \frac{4}{g_RL} e^{-L/g_R}.
\end{equation}
In this large-\(L\) regime, the energy eigenvalue in half-turn space lies deeper than in the three-torus (Eq.(\ref{approximation3Dtori})).

\section{Relevance to Cosmological Scales}\label{sectioniv}

Quantum non-locality implies that cosmic global properties may affect local quantum systems. In Section~\ref{sectioniii}, we showed some examples and proved how topology could induce deviations in the energy eigenvalues of a quantum particle with mass $m$ in a bound state of a Dirac delta potential. This can serve as a preliminary study but exhibits some generic features.

 To this end, we define the relative correction, quantified by the dimensionless ratio
\begin{equation}\label{parametereta}
\eta \equiv \frac{|\tilde{E}| - |\tilde{E}^{(0)}|}{|\tilde{E}^{(0)}|},
\end{equation}
where \( \tilde{E}^{(0)} \) denotes the rescaled energy eigenvalue in the \( \mathbb{R} \) (1D case) and \( \mathbb{R}^3 \) (3D case) cases.
Furthermore, we will compute the values of the scale factor at which these effects begin to become significant, in order to demonstrate the claim made in \citet{Perivolaropoulos:2017rgq} regarding the potential observability of topological effects in the early universe.

There are two ways to define distance in an expanding universe: the comoving distance\footnote{Note that the relationship between space  and its universal covering is identical to that between comoving space and its comoving  
universal covering\citep{Lachieze-Rey:1995qrb}. },  
which remains fixed as the universe expands, and the physical distance, which increases  
due to the expansion.   
The total comoving distance traveled by light  
emitted from an object at time $t$ (when the normalized scale factor was $a(t)=R(t)/R_0$), is given by:  
\begin{equation}\label{chi}
\chi(t)= \frac{c}{R_0}\int_{t}^{t_0} \frac{dt'}{a(t')} = \frac{c}{R_0}\int_{0}^{z} \frac{dz'}{H(z')}.
\end{equation}
where $t_0$ denotes the cosmic time today and $R(t_0)\equiv R_0$.  
The (present) particle horizon is the comoving distance that light could have traveled since $t = 0$ or to an infinite redshift, given by Eq.~(\ref{chi}). In particular, the \emph{particle horizon} defines the boundary between observable and unobservable regions, and its physical distance from us ($\chi = 0$) may be an appropriate choice for a maximum length in quantum mechanics \citep{Perivolaropoulos:2017rgq,Skara:2019uzz}. The physical distance to this maximum observable scale $l_p$ at cosmic time $t$ is given by
\begin{equation}\label{physicaldistance}
    l_p(t)=a(t)\int_{0}^{t} \frac{c\,dt'}{a(t')}=ca(t)\int_{0}^{a(t)} \frac{da'}{a'^2H(a')}
\end{equation}
Note that the definition of the particle horizon in Eq.(\ref{physicaldistance}) applies to both simply and multiply connected Universes, defined in terms of the universal covering space \citep{Lachieze-Rey:1995qrb}.

In Eqs.~\eqref{discreteenergy},~\eqref{enegryeigenstatestorus},~\eqref{1overgrhts}, the length scale \( L \) is interpreted as a physical length. We therefore identify \( L/2 = l_p(a) \) (we will use Eq.~\eqref{physicaldistance}, adopting the Planck 2018 cosmological parameters \( (\Omega_{\rm m0}, \Omega_{\rm r0}, \Omega_{\Lambda 0}) \)~\cite{Planck:2018vyg}).

\subsection{Quantum Bound States in the \(L \gg g_R\) Regime}

\subsubsection{1D case}

As an example to develop intuition, let us examine the one-dimensional potential \( V(x) = -\epsilon \delta(x) \), where the  coupling parameter is defined as \( g \equiv m\epsilon/\hbar^2 > 0 \) (with units of inverse length). This sets the characteristic length scale of the system to \[ \alpha \approx 0.529 \times 10^{-10}\,\text{m}. \]

 To estimate the magnitude of corrections to a bound energy eigenvalue in the present era, we approximate in the large-\(L\) limit using Eq.~\eqref{approximation1d} and for the present scale factor (\( a = 1 \)), Eq.(\ref{parametereta})  yields
\begin{equation*}
   \eta \sim e^{-8.2 \times 10^{36}}.
\end{equation*}
\subsubsection{3D case}

In the large-\(L\) limit, and to leading order, the energy eigenvalues for the 3-Torus and Half-turn space can be expressed in a unified form:
\begin{equation}\label{energyperturbtop}
    |\tilde{E}| \simeq \frac{1}{2g_R^2}\left(1 + C_\Gamma \frac{2g_R}{L}e^{-L/g_R}\right),
\end{equation}
where $C_\Gamma$ is a topology-dependent coefficient: $C_\Gamma = 6$ for the 3-Torus ($E_1$) and $C_\Gamma = 4$ for the Half-turn space ($E_2$). We set the characteristic length scale as
\begin{equation*}
    g_R = \frac{4\pi \epsilon_0 \hbar^2}{m_e e^2} \approx 0.529 \times 10^{-10} \, \text{m}.
\end{equation*}
Evaluating Eq.~(\ref{energyperturbtop})  at scale factor \(a = 1\), one finds that the resulting correction is 
significantly weaker than in the analogous one-dimensional case.  
Although the parameter \(\eta\) carries the same exponential dependence as in 1D, it is further suppressed by the extra factor 
\(g_R/L\sim 10^{-36}\).





\subsection{Quantum Bound States in the \(L \sim g_R\) Regime}

\begin{figure}[h]
\centering
\includegraphics[width=1\columnwidth]{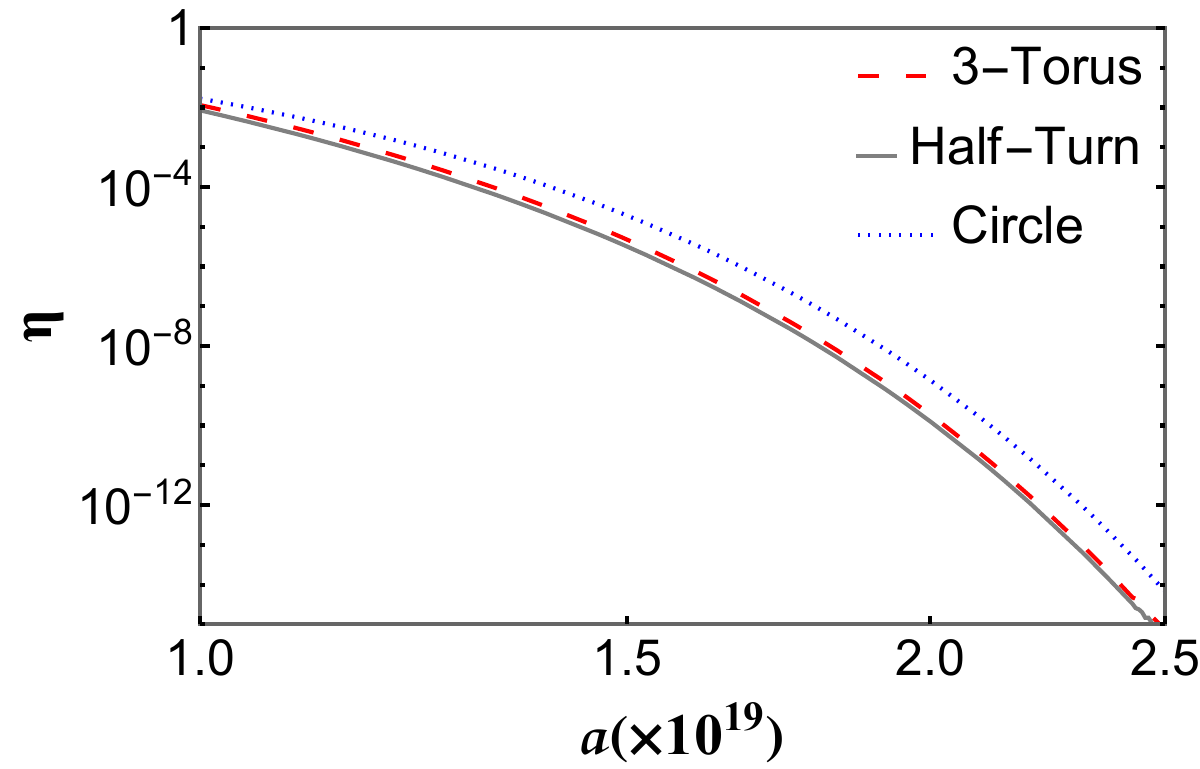}
\caption{We present numerical results for the dimensionless ratio~$\eta$ as a function of the scale factor~$a$  in the $L\sim g_R$ regime. Employing the Planck 2018 cosmological parameters 
\((\Omega_{\rm m0},\Omega_{\rm r0},\Omega_{\Lambda0})\)\citep{Planck:2018vyg}, we compute the physical length  scale
\(L/2=l_p(a)\) via Eq.~\eqref{physicaldistance}. The eigenvalue spectrum is then obtained by numerically solving 
Eqs.~\eqref{discreteenergy}, \eqref{enegryeigenstatestorus} and \eqref{1overgrhts}.}
\label{fig:2}
\end{figure}

Figure~\ref{fig:2} presents the dimensionless parameter $\eta$, defined in Eq.~\eqref{parametereta}, as a function of the scale factor $a$ for the Circle, 3-torus, and half-turn space topologies. The spectrum is obtained numerically: in the case of a one‐dimensional Dirac delta potential by solving 
    Eq.~\eqref{discreteenergy}, and in the case of a three‐dimensional Dirac delta potential by  solving 
    Eqs.~\eqref{enegryeigenstatestorus} and~\eqref{1overgrhts} with a mode cutoff $|n_{i,\rm max}|=20$. Modes with larger $|n_i|$ contribute negligibly due to the exponential suppression $\sim e^{-n\sqrt{2|\tilde E|}\,L}/n$.
   The main features are:\begin{itemize}
 
  \item \textbf{Enhanced binding in compact spaces:} 
    The  3-torus and half-turn (or Circle) space topologies deepen the bound‐state energy \(\tilde{E}=-|\tilde{E}|\) relative to \(\mathbb{R}^3\) (or $\mathbb{R}$). 
     \item \textbf{Topology‐dependent magnitudes:} 
    At a fixed scale factor~\(a\), the 3‐torus topology produces larger energy shifts than the half‐turn space. Furthermore, at the same value of~\(a\), the Circle exhibits even greater energy corrections than the 3‐torus, reflecting its stricter boundary constraints in reduced dimensionality.

  \item \textbf{Significant deviation in energy eigenvalues when $L\sim g_R$:} 
       The correction parameter $\eta$ reaches the percent level once the characteristic length $L$ becomes comparable to the  $g_R$. For example, as shown in Figure~\ref{fig:2}, if $g_R$ is set equal to the Bohr radius, then at  
  $a\sim10^{-19}$, the physical length scale is  
  $l_p(a)\sim10^{-10}\,\mathrm{m}$.
\end{itemize}

Overall,  we observe a deepening of the binding potential for each compact topology, in contrast to the \(\mathbb{R}^3\) case. 
Non‐negligible spectral shifts arise when $L\sim g_R$; in our example, this requires a scale factor $a\lesssim10^{-19}$, corresponding to the electroweak epoch~\citep{Povh:2015gox}, at which point the fundamental length scale is $l_p(a)\sim10^{-10}\,\mathrm{m}$.

\section{Conclusion}

Focusing on cosmic-inspired compact flat topologies ($E_1$-$E_{2}$), we calculate explicit spectral shifts for a quantum particle in a bound state of a 3D Dirac delta potential (see Eqs.~\eqref{enegryeigenstatestorus} and \eqref{1overgrhts}). The energy eigenvalue equations for all cases are derived, revealing distinctive topological signatures in the energy spectrum. For instance, in the large-\(L\) limit, and to leading order, the energy eigenvalues for the 3-Torus and Half-turn space can be expressed in a unified form:
\begin{equation*}
    E \simeq- \frac{\hbar^2}{2mg_R^2}\left(1 + C_\Gamma \frac{2g_R}{L}e^{-L/g_R}\right),
\end{equation*}
where $C_\Gamma$ is a topology-dependent coefficient: $C_\Gamma = 6$ for the 3-Torus ($E_1$) and $C_\Gamma = 4$ for the Half-turn space ($E_2$). In both cases, the compactification deepens  the binding relative to the $\mathbb{R}^3$ result, with $C_\Gamma$ encoding the specific topology dependence.

The global topology of the Universe could, in principle, influence the quantum energy eigenvalues of bound states through boundary‐condition constraints. To investigate this, we model a quantum particle in a three‐dimensional Dirac delta potential located at the center of a fundamental cube of side length \(L\), identifying \(L/2 = l_{p}\), as a toy model.
  At the present epoch ($a=1$), measurements of atomic energy eigenvalues cannot detect such effects due to the large length scale \( l_p \sim 10^{26} \, \mathrm{m} \). Detectable shifts could occur, but this would require a scale factor range \( a \lesssim 10^{-19} \), at which the length scale \( l_p \) would be on the order of \( 10^{-10} \, \mathrm{m} \), comparable to the characteristic size of an atom.
 Although direct observational tests in today’s universe (\,\(a=1\)\,) remain beyond reach, our analysis shows that nontrivial cosmic topology can imprint  signatures on quantum spectra. Therefore, theoretical treatments of the very early Universe may need to incorporate potential topological effects to comprehensively capture all possible observational imprints.

Potential extensions of this work include applying the three‐dimensional Dirac‐delta toy model to quarter‐turn space or other cosmic‐inspired flat topologies~\cite{Riazuelo:2003ud,COMPACT:2022gbl}, and performing a systematic review of all flat cosmic‐inspired topologies, relating each to its $C_{\Gamma}$ coefficient in the large‐$L$ limit ($L\gg g_{R}$). Additionally, one could explore the hydrogen atom or quantum fields in these topological settings. Finally, another promising direction involves studying the topologies of curved spaces with \(k=\pm1\)~\cite{Lachieze-Rey:1995qrb,Levin:2001fg,Altunkaynak:2006ik}.

\section*{Data Availability Statement}

This study is based on theoretical and symbolic calculations, with numerical computations performed using Mathematica where applicable. All relevant results and figures are included in the manuscript, and no external datasets were generated or used.

\begin{acknowledgments}
 This research was supported by COST Action CA21136 - Addressing observational tensions in cosmology with systematics and fundamental physics (CosmoVerse), supported by COST (European Cooperation in Science and Technology). 
\end{acknowledgments}

\appendix

\section{The 1D Dirac Delta Potential}\label{App}

In one dimension, the time-independent Schrödinger equation for a particle of mass $m$ is given by  
\begin{equation}\label{sch1D}
    -\frac{\hbar^2}{2m} \frac{\rm d^2}{\rm dx^2} \Psi(x) + V(x) \Psi(x) = E \Psi(x).
\end{equation}
We consider the potential $V(x) = -\epsilon\delta(x)$, define the positive parameter $g \equiv \frac{m\epsilon}{\hbar^2}$, and introduce the characteristic length scale $\alpha \equiv 1/g$. We also denote as  $\tilde{E} \equiv \frac{m E}{\hbar^2}$, which has the scale of inverse length squared.

We obtain the normalized wavefunction corresponding to the bound state as:
\begin{equation}
    \Psi(x)=\left(2 |\tilde{E}|\right)^{\frac{1}{4}}e^{-\left(2 |\tilde{E}|\right)^{\frac{1}{2}}|x|}
\end{equation}
while the parameter $|\tilde{E}|$ attains the value
\begin{equation}\label{energyR}
  |\tilde{E}|=\frac{1}{2\alpha^2}\implies E=-\frac{\hbar^2}{2m\alpha}. 
\end{equation}
Note that by taking the Fourier transform of Eq.~\eqref{sch1D}, Eq.~\eqref{energyR} can be obtained.

Next, we solve Eq.~\eqref{sch1D} on a circular domain, where the spatial coordinate $x \in \left[-\frac{L}{2}, \frac{L}{2}\right]$ is parameterized as $x = R\theta$, with radius $R = \frac{L}{2\pi}$. The system obeys periodic boundary conditions:
\begin{equation}
    \Psi\left(-\tfrac{L}{2}\right) = \Psi\left(\tfrac{L}{2}\right) \quad \text{and} \quad \Psi'\left(-\tfrac{L}{2}\right) = \Psi'\left(\tfrac{L}{2}\right),
\end{equation}
along with wavefunction continuity at $x = 0$.  For the bound state with energy \( \tilde{E} = -|\tilde{E}| \), the wavefunction takes the form
\begin{equation}\label{wavefunctioncircle}
   \Psi(x) = B \cosh\left(\sqrt{2|\tilde{E}|}\left(\tfrac{L}{2} - |x|\right)\right),
\end{equation}
where \( B \) is a normalization constant, which is determined to be
\begin{equation}\label{normalization}
    B = \frac{2}{\sqrt{2L + \sqrt{\frac{2}{|\tilde{E}|}} \sinh\left(\sqrt{2|\tilde{E}|}L\right)}}.
\end{equation}

Finally, the derivative boundary condition at \( x = 0 \) is given by
\begin{equation}\label{boundaryconditiondelta}
    \Psi'_+(0) - \Psi'_-(0) = -2g\Psi(0).
\end{equation}
Substituting \eqref{wavefunctioncircle} into \eqref{boundaryconditiondelta} yields the  condition for the bound state energy:
\begin{equation}\label{energycondition}
      \sqrt{2|\tilde{E}|}=g \coth\left( \frac{\sqrt{|\tilde{E}|}L}{\sqrt{2}} \right),
\end{equation}
which is identical to Eq.(\ref{discreteenergy}).
\section{Some Proofs}\label{AppB}
\begin{itemize}
    \item \textbf{Lemma I:} In the 3-torus space, the Dirac comb satisfies
\begin{equation}\label{torusappb}
    \sum_{\mathbf{n}}\delta(\mathbf{x}-\mathbf{n}L)=\frac{1}{L^3}\sum_{\mathbf{n}}e^{\frac{2\pi i}{L}\mathbf{n}\cdot\mathbf{x}}\quad \text{where} \quad n_i\in \mathbb{Z}.
\end{equation}

\textbf{Proof:} Let us define $\Delta(\mathbf{x}) \equiv \sum_{\mathbf{m}} \delta^{(3)}(\mathbf{x} - \mathbf{m}L)$, where $m_i\in \mathbb{Z}$. Then, from Eqs.~(\ref{eq:fourier_expansion1}-\ref{fourrierhalf2}), we obtain 
\begin{equation*}
    \tilde{\Delta}_{\mathbf{n}}= \int_{[-\frac{L}{2},\frac{L}{2}]^3} \left[\sum_{\mathbf{m}} \delta^{(3)}(\mathbf{x} - \mathbf{m}L)\right] \xi_{\mathbf{n}}^*(\mathbf{x})\,d^3x=
\end{equation*}
\begin{equation*}
    =\int_{[-\frac{L}{2},\frac{L}{2}]^3} \delta^{(3)}(\mathbf{x} ) \xi_{\mathbf{n}}^*(\mathbf{x})\,d^3x+
\end{equation*}
\begin{equation*}
    +\int_{[-\frac{L}{2},\frac{L}{2}]^3} \left[\sum_{\mathbf{m}\neq\mathbf{0}} \delta^{(3)}(\mathbf{x} - \mathbf{m}L)\right] \xi_{\mathbf{n}}^*(\mathbf{x})\,d^3x=\frac{1}{L^{3/2}},
\end{equation*}
where the second term vanishes since $x_i \in [-L/2, L/2]$, and we have used the fact that the basis functions are given by Eq.~(\ref{basistorus}).
 Then by using Eq.(\ref{fourrierhalf2}) we get
\begin{equation*}
    \Delta(\mathbf{x})=\sum_{\mathbf{k}}\tilde{\Delta}_{\mathbf{k}}\xi_{\mathbf{k}}(\mathbf{x})=\frac{1}{L^3}\sum_{\mathbf{n}}e^{\frac{2\pi i}{L}\mathbf{n}\cdot\mathbf{x}}.\quad \Box
\end{equation*}

\item Let us define the sets:
\begin{align*}
    \mathcal{I}_0 &= \left\{ \mathbf{n} \in \mathbb{Z}^3 \;\middle|\; \mathbf{n} = (0, 0, n_z),\; n_z \in 2\mathbb{Z} \right\}, \\
   \mathcal{I}^* &= \left\{ \mathbf{n} \in \mathbb{Z}^3 \;\middle|\; 
    \begin{array}{l}
        n_x \in \mathbb{Z}^+, \; n_y \in \mathbb{Z},\, n_z\in2\mathbb{Z}\\[0.5ex]
        \text{or } n_x = 0, \; n_y \in \mathbb{Z}^+, \; n_z \in 2\mathbb{Z}
    \end{array}
    \right\}.
\end{align*}
We then define the function $\Delta(\mathbf{x})$ as:
\begin{equation}\label{diraccomb}
    \Delta(\mathbf{x}) \equiv  \sum_{\mathbf{n} \in \mathcal{I}_0\cup\mathcal{I}^*} \delta^{(3)}(\mathbf{x} - \mathbf{n}L).
\end{equation}
\textbf{Lemma II:} In the Half-Turn space, the Eq.(\ref{diraccomb}) satisfies
\begin{widetext}
    \begin{equation}\label{halfturndiracdcombAppB}
    \Delta(\mathbf{x}) =  \frac{1}{L^3}\sum_{\mathbf{n}\in \mathcal{I}_0}e^{i\frac{2\pi}{L}\mathbf{n}\cdot\mathbf{x}}+\frac{1}{L^3}\sum_{\mathbf{n}\in \mathcal{I}^*}e^{i\frac{2\pi n_z}{L}z}\left[e^{i \frac{2\pi}{L}(n_xx+n_y y)}+e^{-i \frac{2\pi}{L}(n_xx+n_y y)}\right],
\end{equation}
\end{widetext}

\textbf{Proof:}
From Eqs.(\ref{eq:fourier_expansion1}) and (\ref{basishalf}), we expand as 
\begin{equation}\label{expansionproof}
    \Delta(\mathbf{x})=\sum_{\mathbf{n}\in \mathcal{I}_0\cup\mathcal{I}^*}\tilde{\Delta}_{\mathbf{n}}\xi_{\mathbf{n}}(\mathbf{x}).
 \end{equation}
The Fourier coefficients \(\tilde{\Delta}_{\mathbf{n}}\) are obtained from Eq.~\eqref{fourrierhalf2}:
\begin{equation*}
    \tilde{\Delta}_{\mathbf{n}}= \int_{[-\frac{L}{2},\frac{L}{2}]^3} \left[\sum_{\mathbf{m}\in\mathcal{I}_0\cup\mathcal{I}^*} \delta^{(3)}(\mathbf{x} - \mathbf{m}L)\right] \xi_{\mathbf{n}}^*(\mathbf{x})\,d^3x=\xi_{\mathbf{n}}^*(\mathbf{0}),
\end{equation*}
since the only contributing term to \(\tilde{\Delta}_{\mathbf{n}}\) is the one with \(\mathbf{m} = \mathbf{0}\).
 Given that we can write
\[
\sum_{\mathbf{n} \in \mathcal{I}_0 \cup \mathcal{I}^*} \tilde{\Delta}_{\mathbf{n}} \, \xi_{\mathbf{n}}(\mathbf{x})
= \sum_{\mathbf{n} \in \mathcal{I}_0} \tilde{\Delta}_{\mathbf{n}} \, \xi_{\mathbf{n}}(\mathbf{x}) 
+ \sum_{\mathbf{n} \in \mathcal{I}^*} \tilde{\Delta}_{\mathbf{n}}\, \xi_{\mathbf{n}}(\mathbf{x}),
\]
then Eq.~\eqref{expansionproof} can be rewritten as
\begin{equation*}
    \Delta(\mathbf{x})=\sum_{\mathbf{n}\in \mathcal{I}_0}\xi_{\mathbf{n}}^*(\mathbf{0})\xi_{\mathbf{n}}(\mathbf{x})+\sum_{\mathbf{n}\in \mathcal{I}^*}\xi_{\mathbf{n}}^*(\mathbf{0})\xi_{\mathbf{n}}(\mathbf{x}),
 \end{equation*}

 which yields Eq.(\ref{halfturndiracdcombAppB}). $\Box$
\end{itemize}

    


\newpage
\bibliography{main}

\end{document}